\documentstyle[11pt,aaspp4]{article}

\lefthead{Hogerheijde et al.}
\righthead{Sub-arcsecond imaging at 267~GHz of a young binary system}
\begin{document}



\title{Sub-arcsecond imaging at 267~GHz of a young binary system:
Detection of a ${\bf <70}$~AU radius dust disk around T~Tau~N}

\author{Michiel R. Hogerheijde\altaffilmark{1}, Huib Jan van
Langevelde\altaffilmark{2}, Lee G. Mundy\altaffilmark{3}, Geoffrey
A. Blake\altaffilmark{4}, and Ewine F. van Dishoeck\altaffilmark{1}}

\altaffiltext{1}{Sterrewacht Leiden, P.O. Box 9513, 2300 RA, Leiden, The
Netherlands}
\altaffiltext{2}{Joint Institute for VLBI in Europe, P.O. Box 2, 7990 AA,
Dwingeloo, The Netherlands}
\altaffiltext{3}{Department of Astronomy, University of Maryland, College Park,
MD~20742}
\altaffiltext{4}{Division of Geological and Planetary Sciences, California
Institute of Technology, MS 150--21, Pasadena, CA~91125}



\begin{abstract}

The young binary system T~Tau was observed with the Owens Valley
Millimeter Array in 267~GHz continuum and HCO$^+$ $J$=3--2 emission at
0{\farcs}8 resolution, with the JCMT--CSO single-baseline
interferometer in 357~GHz continuum, and with the W.~M.~Keck telescope
at $\lambda$=4 $\mu$m.  The 267~GHz emission is unresolved with a flux
of $397\pm 35$~mJy, located close to the position of the optical star
T~Tau~N. An upper limit of 100~mJy is obtained toward the infrared
companion T~Tau~S.  The 357~GHz continuum emission is unresolved, with
a flux of $1.35\pm 0.68$~Jy. HCO$^+$ $J$=3--2 was detected from a
2$''$ diameter core surrounding T~Tau~N and S. Both stars are detected
at 4 $\mu$m, but there is no evidence of the radio source T~Tau~R.

We propose a model in which T~Tau~S is intrinsically similar to
T~Tau~N, but is obscured by the outer parts of T~Tau~N's disk. A fit
to the spectral energy distribution (SED) between 21~cm and
1.22~$\mu$m is constructed on this basis. Adopting a r$^{-1}$ surface
density distribution and an exponentially truncated edge, disk masses
of $0.04\pm 0.01$ $M_\odot$ and $6\times 10^{-5} - 3\times
10^{-3}$~$M_\odot$ are inferred for T~Tau~N and T~Tau~S, respectively.
A 0.005--0.03~$M_\odot$ circumbinary envelope is also required to fit
the millimeter to mid-infrared SED.

\end{abstract}

\keywords{ISM: molecules --- stars: formation --- stars: low mass,
brown dwarfs --- stars: pre--main sequence --- binaries: close}



\section{Introduction}

Many low-mass stars form in multiple systems (cf.~Ghez et~al.\
\markcite{ghe93}1993), either through capture, fragmentation of the
collapsing core, or condensation from the primary's disk (e.g.,
Bodenheimer, Ruzmaikina \& Mathieu \markcite{bod93}1993). The presence
of a close companion must influence the evolution of the star and the
possible development of an accretion disk (e.g., Jensen et~al.\
\markcite{jen96b}1996). Observations of the distribution of the bulk
of the gas and dust in young and forming multiple systems are needed
to better understand their evolution, which may be very different from
that of a single star. This Letter presents high resolution continuum
and spectral line observations of one such close multiple system:
T~Tauri.

A prototypical low-mass young stellar object (YSO), T~Tau is a
0{\farcs}7 separation binary (projected distance of 100~AU at 140~pc),
consisting of an optically visible star, T~Tau~N, and an infrared
companion, T~Tau~S (Dyck, Simon \& Zuckerman
\markcite{dyc82}1982). Ray et~al.\ \markcite{ray97}(1997) report the
presence at $\lambda=6$ cm of a third source, possibly a star,
T~Tau~R. The system is surrounded by 0.03 to 0.3~$M_\odot$ of gas and
dust distributed on 3000 to $10^4$~AU scales (e.g., Schuster, Harris
\& Russell \markcite{sch97}1997; Hogerheijde et~al.\
\markcite{mrh97}1997). Both stars appear to drive outflows, at least
one of which is directed close to the line of sight (Edwards \& Snell
\markcite{edw82}1982, Beckwith et~al.\ \markcite{bec78}1978;
van~Langevelde et~al.\ \markcite{hvl94b}1994; Herbst et~al.\
\markcite{her97}1997).  Stellar light curves indicate that the
rotation axis of T~Tau~N is inclined by only $\sim 19^\circ$ (Herbst
et~al.\ \markcite{her97}1997). From the strength and variability of
its free-free radio emission, T~Tau~S appears to be a more active
outflow source than T~Tau~N (Skinner \& Brown \markcite{ski94}1994;
Ray et~al.\ \markcite{ray97}1997).

The infrared nature and activity of T~Tau~S have been explained in a
number of ways: through obscuration by a remnant protostellar envelope
surrounding T~Tau~S (Dyck et~al.\ \markcite{dyc82}1982) or around the
binary system (Calvet et~al.\ \markcite{cal94}1994), or through
obscuration caused by periodic perturbation of circumstellar material
in the binary orbit (Koresko, Herbst \& Leinert \markcite{kor97}1997).
Until recently, the T~Tau binary has only been resolved at centimeter
and near-infrared wavelengths. All information on the gas and dust in
the system has been derived from (sub-)millimeter and far-infrared
measurements, which did not separate the pair.  In this Letter we
present sub-arcsecond resolution 267 and 357~GHz continuum
observations showing that the compact dust emission from the system is
confined to the surroundings of T~Tau~N (see also van~Langevelde
et~al.\ \markcite{hvl97}1997). In addition, 0{\farcs}4 resolution
imaging confirms that T~Tau~S is brighter than T~Tau~N at
$\lambda$=4~$\mu$m.  The spatially resolved (sub-)millimeter fluxes,
and literature values for centimeter and infrared observations, allow
a fit to the spectral energy distributions (SEDs) of the individual
sources, constraining the mass and size of the disk around T~Tau~N and
the circumstellar material around T~Tau~S.

\section{Observations}

Continuum emission at 267~GHz and HCO$^+$ $J$=3--2 line emission were
imaged with the Millimeter Array at the Owens Valley Radio Observatory
(OVRO) on 1994, December~1. Baselines between 20 and 200~k$\lambda$
were sampled, resulting in a $0{\farcs}77\times 0{\farcs}96$
synthesized beam for a robust weighting of +1. The continuum data were
integrated over a 1~GHz band width, yielding a $1\sigma$ RMS noise
level of 30~${\rm mJy\,bm^{-1}}$. The HCO$^+$ 3--2 data were recorded
in a 64 channel correlator with a resolution of 0.5~MHz or 0.56~${\rm
km\,s^{-1}}$; the $1\sigma$ RMS noise in a channel is 1.3~${\rm
Jy\,bm^{-1}}$, or 19~${\rm K\,km\, s^{-1}}$ for natural weighting. The
antenna-based complex gains of the OVRO instrument were derived with
the MMA package, using PKS~0528+134 as the phase calibrator and
3C~454.3 as the flux calibrator, for which a flux of 6.50~Jy was
derived from observations of planets. The data were mapped and
analyzed with the MIRIAD software package.  Like the radio positions
of T~Tau~N and S, the coordinates of the 267~GHz emission are tied to
the radio reference frame; the estimated uncertainty in the absolute
positions of 0{\farcs}2 is dominated by the accuracy of the array
baselines.

The 357~GHz continuum emission was observed together with HCO$^+$ 4--3
line emission using the single-baseline interferometer (SBI) of the
James Clerk Maxwell Telescope (JCMT) and the Caltech Submillimeter
Observatory (CSO), on 1994, October~28. Projected baselines ranged
between 140 and 195~k$\lambda$, resulting in an effective resolution
of $\sim 0{\farcs}7$.  The data were recorded in a 500~MHz wide band
with 400 channels, and vector-averaged over 100~second intervals. The
gain of the instrument was $\sim 135$~${\rm Jy\,K^{-1}}$, derived from
PKS~0528+134. Since the phase variations on the JCMT--CSO baseline
cannot be tracked, only the 100~s vector-averaged visibility
amplitudes are used (cf.~Lay et~al.\ \markcite{lay94}1994).

Imaging at $\lambda$=4~$\mu$m was acquired with the W.~M.~Keck
Telescope on 1997, February~7, using the Near Infrared Camera and
Br$\alpha$--continuum filter (3.97--4.02~$\mu$m).  The detector array
has 0{\farcs}15 pixels. The final image contains 2000 integrations of
45~milliseconds each.  Atmospheric conditions were good, with a seeing
of 0{\farcs}3--0{\farcs}4. The image was processed with the IRAF
package.

\section{Results}

The 267~GHz continuum emission, shown in Fig.~1a, is unresolved with a
peak flux of $369\pm 30$~${\rm mJy\,bm^{-1}}$. A point-source fit to
the visibilities yields a flux of $397\pm 35$~mJy, and a position
within 1$\sigma$ (0{\farcs}2) of both T~Tau~N and T~Tau~R
(Table~1). The $3\sigma$ upper limit on the flux toward T~Tau~S is
100~mJy. Although T~Tau~R, at 0{\farcs}13, is slightly closer to the
continuum peak than T~Tau~N (0{\farcs}20), we attribute the 267~GHz
emission to T~Tau~N, due to the uncertain nature of T~Tau~R.  Based on
the compactness of the source and the emission at infrared and
sub-millimeter wavelengths (Herbst et~al. \markcite{her97}1997;
Beckwith et~al. \markcite{bec90}1990), we interpret the 267 GHz
emission as arising from a disk of radius $<0{\farcs 45} \approx
70$~AU around the optical star.

The $\lambda$=4~$\mu$m Keck image of the T~Tau system is shown in
Fig.~1b in grey scale, with contours of the 267~GHz emission
superposed.  The binary is clearly resolved and is consistent with two
individual point sources. The 4~$\mu$m flux from T~Tau~S is 1.5 times
that from T~Tau~N.  No absolute positional information is available
for the Keck data, so the image has been aligned with the radio
positions of T~Tau. The binary separation at 4~$\mu$m is 0{\farcs}70,
consistent with measurements at centimeter wavelengths. No evidence is
found for emission from T~Tau~R.

The 357~GHz continuum SBI data are consistent with emission from a
single point source and yield a vector-averaged flux of $1350\pm 675$
mJy, where the uncertainty is dominated by the flux calibration.  The
observed emission at 267 and 357~GHz accounts for 60\%--80\% of the
single-dish fluxes at these wavelengths on $19''$ and $13''$ scales,
respectively (Moriarty-Schieven et~al.\
\markcite{mor94}1994). Including the 2.7~mm point-source flux from
Hogerheijde et~al.\ \markcite{mrh97}(1997), a spectral index of
$2.6\pm 0.2$ is found for the compact emission.

Emission in the HCO$^+$ 3--2 line is detected only on baselines
$<70$~k$\lambda$, and over 3--11~${\rm km\,s^{-1}}$. The naturally
weighted image (Fig.~1c) shows a $2''$ core, centered between T~Tau~N
and S, with a peak flux of $6.1\pm 0.4$~${\rm Jy\,bm^{-1}}$ or
745~${\rm K\,km\, s^{-1}}$.  This is 25\% of the $19''$ single-dish
line flux, comparable to the 10\% recovered at HCO$^+$ 1--0 by OVRO
(Hogerheijde et~al.\ \markcite{mrh98b}1998).  No emission from HCO$^+$
4--3 was detected in the SBI data with a statistical upper limit of
$\sim 0.35$~Jy, or $\sim 7$~K, for a $1''$ source size, consistent
with the non-detection of HCO$^+$ 3--2 emission on baselines
$>70$~k$\lambda$.

\section{Understanding the nature of T~Tau~S and T~Tau~N}

Our observations of 267~GHz continuum emission show that most of the
material on 100~AU scales is associated with T~Tau~N, most likely in a
disk. Although the near-infrared SED of T~Tau~S also indicates the
presence of a disk, the 267~GHz upper limit strictly constrains the
mass.  Bate \& Bonnell \markcite{bat97}(1997) found that material
infalling toward a binary system preferably forms a disk around the
more massive primary, which our observations show to be T~Tau~N. How
does the difference in disk mass for T~Tau~N and S impact our
understanding of the T~Tau system, especially the obscured, infrared
nature of T~Tau~S?

Of the many scenarios which have been proposed to explain the infrared
nature of T~Tau~S (see \S 1), that of dynamical interaction is
particularly interesting because it links the obscuration of infrared
companions in binary systems with their observed variability (Koresko
et~al.\ \markcite{kor97}1997). Based on our new data for T~Tau, we
propose a variation on this model, wherein T~Tau~S is obscured
directly by the disk around T~Tau~N. This requires the disk to extend
to $\gtrsim 100$~AU, at a modest inclination with respect to the
binary orbital plane. For the assumed dust properties (Ossenkopf \&
Henning \markcite{oss94}1994; their model `thin 6'), a surface density
of 0.12~${\rm g\,cm^{-2}}$ is required to obscure an assumed stellar
photosphere for T~Tau~S of $\sim 5000$~K.

To further investigate this model and derive mass estimates for the
disks around T~Tau~N and S, and for the circumbinary envelope, a fit
to the SED between 21~cm and 1.22~$\mu$m is constructed (Fig.~2). The
model for T~Tau~N is based on a stellar photosphere of 5250~K (K0
star), and a temperature distribution in the disk characterized by a
radial power law with index $-q$ with an intrinsic luminosity $L_{\rm
d}$ (Adams, Lada \& Shu \markcite{ada87}1987; eq.~[A22]). From the
optically thick 1.22 to 20 $\mu$m range, estimates are obtained of the
stellar radius, $R_*=3.4$ $R_\odot$, the inner radius of the disk,
$R_{\rm in}=5.1$ $R_\odot$, the disk luminosity, $L_{\rm d}=3.0$
$L_\odot$, and the index, $q=0.51$ (cf.\ Herbst et~al.\
\markcite{her97}1997).

At longer wavelengths, where the disk is largely optically thin, the
surface density is important. The density distribution depends on
angular momentum transport processes like viscosity and disk wind
(cf.\ Adams \& Lin \markcite{ada93}1993). The gravitational
interaction with the binary companion is expected to truncate the disk
at some radius (cf.\ Lin \& Papaloizou \markcite{lin93}1993). The
surface density is approximated here by a radial power law $\Sigma
\propto r^{-1}$, which is truncated at radii $r>R_{\rm trunc}$ by an
exponential taper, $\exp [-(r-R_{\rm trunc})/r_e]$. It is found that
this exponential fall-off is required to permit the low surface
density of 0.12 g~cm$^{-2}$ needed for the obscuration of T~Tau~S
while and still fitting the flux of T~Tau~N. The exact functional form
of the truncation or the adopted outer radius is not important for the
derived masses for $R_{\rm trunc} \gtrsim 50$ AU. Assuming $R_{\rm
trunc} = 70$ AU, the only free model parameter left is the ${1\over
e}$ radius $r_e$. The 267~GHz flux constrains $r_e$ to 7--8 AU, with a
corresponding disk mass of $0.04\pm 0.01$ $M_\odot$.

The SED of T~Tau~S is fitted with the same model and parameters, plus
obscuration by the disk of T~Tau~N. The only differences are the disk
mass, along with the luminosity of 15~$L_\odot$ and index of q=0.54
required to fit the higher infrared fluxes. An interstellar extinction
of A$_{\rm V}$=1.44~mag is applied to the whole system (Cohen \& Kuhi
\markcite{coh79}1979), and free-free spectral indices of 0.6 and 0.2
are used for T~Tau~N and S, respectively. The assumed stellar
temperature is unimportant since the disk luminosity dominates the SED
at $\lambda > 2$ $\mu$m. The disk mass is constrained by the 20~$\mu$m
flux and the 267~GHz upper limit to lie between $6\times 10^{-5} -
3\times 10^{-3}$~$M_\odot$.

To fit the unresolved (sub-)millimeter and infrared observations, a
circumbinary envelope of 0.005--0.03~$M_\odot$ is also required,
depending on the disk mass of T~Tau~S. This envelope is modeled with a
power law for the density with slope $p=-1.5$, and for the
temperature with slope $q=-0.4$. At the inner radius of 100~AU, a
temperature of 63~K provides a good fit to the peak of the SED. An
outer radius of 3000~AU is used, corresponding to the typical beam
size of the (sub-)millimeter observations. The presence of this
power-law envelope is consistent with the non-detection of extended
267~GHz continuum emission on 16--200 k$\lambda$ baselines at the
obtained noise level. The HCO$^+$ 3--2 emission on 16--70 k$\lambda$
baselines can be explained by the sensitivity of this line to density,
which results in a more peaked brightness distribution, although
additional emission from, e.g., material in the walls of the outflow
cavity (cf.\ Hogerheijde et~al.\ \markcite{mrh98b}1998) is required to
fit its absolute flux.

The non-detection of HCO$^+$ 3--2 on baselines $>70$~k$\lambda$ with
an upper limit of $\sim 1.4$~${\rm Jy\,bm^{-1}}$, or $\sim 171$~${\rm
K\,km\, s^{-1}}$, sets a rough upper limit to the HCO$^+$ abundance in
the disk. Assuming an excitation temperature of 60~K and neglecting
any optical depth effects, an upper limit to the abundance of
$1\times 10^{-8}$ is found.

The limit placed on the amount of circumstellar material around
T~Tau~S by our 267~GHz measurements places constraints on the nature
of the obscuration.  While a small edge-on disk around T~Tau~S could
provide the observed extinction, the disk around T~Tau~N is a more
likely candidate. In either case, it is implied that the binary orbit
and the respective disks are not coplanar. A circumstellar envelope
around T~Tau~S alone is unlikely as obscuring agent, since it implies
that either the current projected binary separation is a chance close
projection, or that the envelope is $<50$~AU and hence dynamically
short-lived. An envelope around both T~Tau~N and S requires a special
geometry which allows T~Tau~N to suffer little extinction while
heavily obscuring T~Tau~S.

To test the hypothesis that T~Tau~N's disk is obscuring T~Tau~S,
sub-arcsecond resolution observations at sub-millimeter wavelengths
and at 10--20~$\mu$m are needed to place firmer limits on the
distribution of the circumstellar material around T~Tau~S. For
example, monitoring of the near-infrared variability and the shape of
the 10~$\mu$m silicate absorption could distinguish intrinsic
variations in T~Tau~S from inhomogeneities in the obscuring disk or
changes in the star--disk geometry. Observation of the 20--200~$\mu$m
spectrum planned for the {\it Infrared Space Observatory\/} will help
to further constrain the SED model, since the emission from the disks
and the envelope peak at these wavelengths. Some of the other infrared
companions to T~Tauri~stars might be explained by disk obscuration as
well (e.g., Haro~6--10 at 1{\farcs}2 separation, Leinert \& Haas
\markcite{lei89}1989; XZ~Tau at 0{\farcs}3, Haas et al.\
\markcite{haa90}1990; UY~Aur at 0{\farcs}9, Herbst et al.\
\markcite{her95}1995). Millimeter observations at sub-arcsecond
resolution are needed to determine the material distributions in these
systems.

\acknowledgments

The authors wish to thank O.~Lay for a critical reading of the
manuscript, and him and J.~Carlstrom for assistance in the observation
and reduction of the SBI data. The telescope staffs are thanked for
support during the observations. E.~v.~D. acknowledges support by
NWO/NFRA, G.~A.~B. by NASA (NAGW--2297, NAGW--1955), L.~G.~M. by NASA
(NAG 5--4429), H.~J.~v.~L. by the European Union (CHGECT920011). The
OVRO Millimeter Array and the CSO are operated by the Caltech under
funding from the NSF (AST96--13717, AST93--13929). The JCMT is
operated by JAC on behalf of PPARC UK, NWO Netherlands, and NRC
Canada. The W.~M.~Keck Observatory is operated as a scientific
partnership between Caltech, Univ.~of California, and NASA. It was
made possible by the generous financial support of the W.~M.~Keck
Foundation.








\newpage

\figcaption[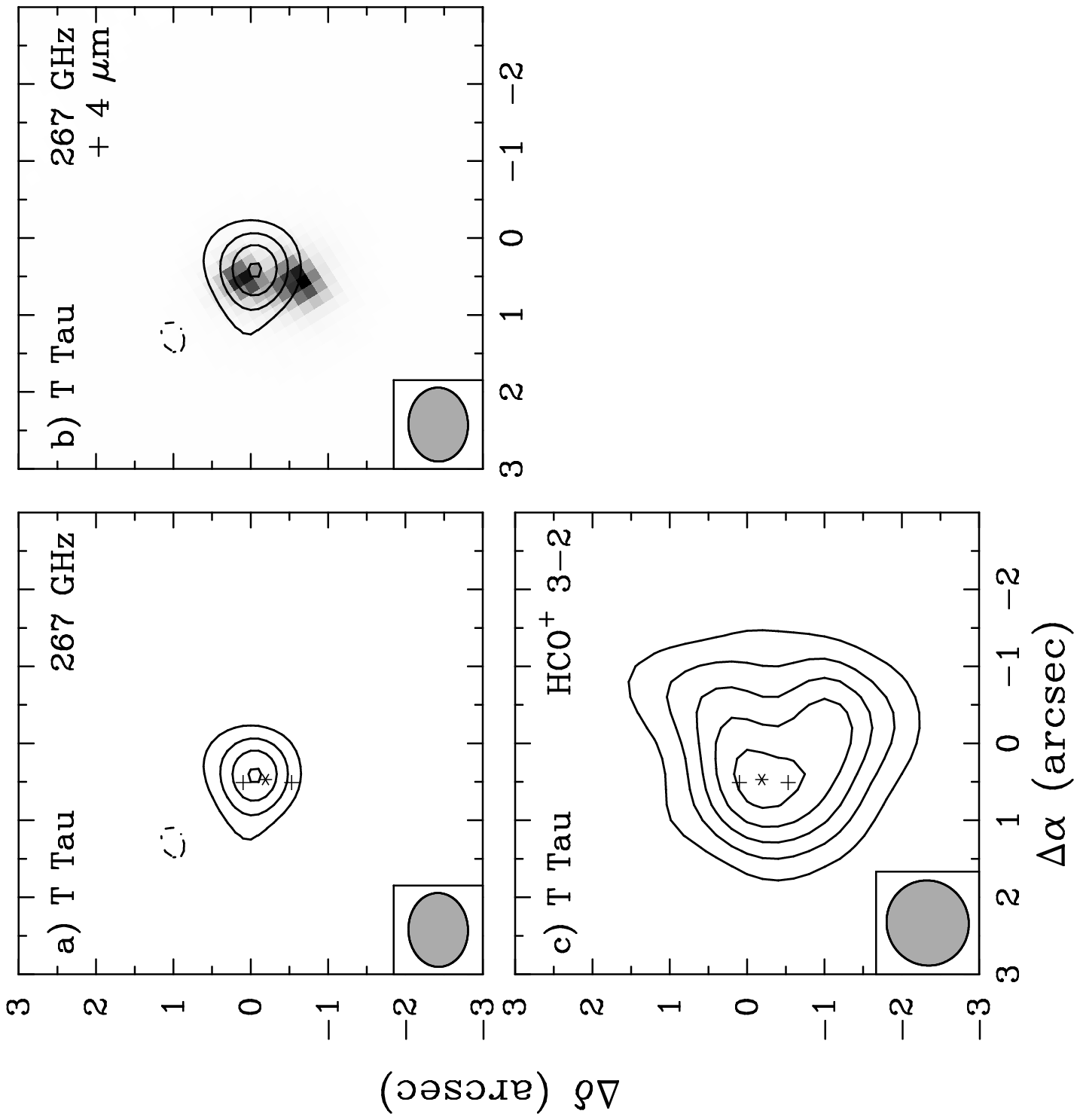] 
{({\it a\/}) Cleaned image of 267~GHz continuum emission of T~Tau using robust
weighting of +1. Contours are drawn at 3$\sigma$ intervals of 90~${\rm
mJy\,bm^{-1}}$. T~Tau~N and S are indicated by crosses, T~Tau~R by an
asterisk. ({\it b\/}) Same 267~GHz contours overlaid on the 4~$\mu$m image
(grey scale), which has been aligned with the radio positions of
T~Tau~N and S. ({\it c\/}) Velocity integrated HCO$^+$ 3--2 emission on
baselines $<120$~k$\lambda$. Contours are drawn at 3$\sigma$ intervals
of 9.2~${\rm Jy\,bm^{-1}\,km\,s^{-1}}$ (134~${\rm K\,km\, s^{-1}}$).} 

\figcaption[fig2.eps] 
{Spectral energy distribution of T~Tau~N ({\it top panel\/}) and S
({\it lower panel\/}), and model curves for T~Tau~S $M_{\rm d}=5\times
10^{-4}$~$M_\odot$, and envelope $M=0.02$~$M_\odot$.  The data points
include our 267~GHz and 357~GHz measurements and literature data
(Beckwith et~al. 1990; Beckwith \& Sargent 1991; Herbst et~al.\ 1997;
Hogerheijde et~al.\ 1997; Ray et~al.\ 1997; Skinner \& Brown 1994;
Weintraub, Sandell \& Duncan 1989; Weintraub, Masson, \& Zuckerman
1987). The model is discussed in the text.}



\newpage

\begin{deluxetable}{lrrr}
\tablecolumns{4}
\tablewidth{0pt}
\tablecaption{Source positions and 267 GHz fluxes\tablenotemark{a}}
\tablehead{
 & \colhead{$\alpha(1950.0)$} & \colhead{$\delta(1950.0)$} & 
\colhead{$F_\nu$} \nl
\colhead{Source} & \colhead{(hh mm ss)} & \colhead{($^\circ$ $'$ $''$)}
 & \colhead{(Jy)}
}
\startdata
T~Tau~N & 04 19 04.24 & +19 25 04.92 & $0.397 \pm 0.035$ \nl
T~Tau~S & \nodata & \nodata & $<0.10$ \nl
HCO$^+$ 3--2 & 04.19 04.23 & +19 25 04.46 & $29.5\pm 2.8$ \nl
\cutinhead{Radio positions}
T~Tau~N\tablenotemark{b} & 04 19 04.245 & +19 25 05.20 \nl
T~Tau~S\tablenotemark{b} & 04.19 04.250 & +19 25 04.58 \nl
T~Tau~R\tablenotemark{c} & 04.19 04.243 & +19 25 04.88 \nl
\enddata
\tablenotetext{a}{Absolute positional accuracy at 267~GHz $\sim 0{\farcs}2$.}
\tablenotetext{b}{3.6 cm coordinates from Skinner \& Brown 1994.}
\tablenotetext{c}{6.0 cm coordinates from Ray et~al.\ 1997.}
\end{deluxetable}



\input psfig

\null\rightline{\underline{\tt Fig. 1}}
\centerline{\vbox{
\psfig{figure=fig1.ps,height=15truecm,rheight=10truecm,angle=-90}
}}
\vfill\eject
\null\rightline{\underline{\tt Fig. 2}}
\centerline{\vbox{
\psfig{figure=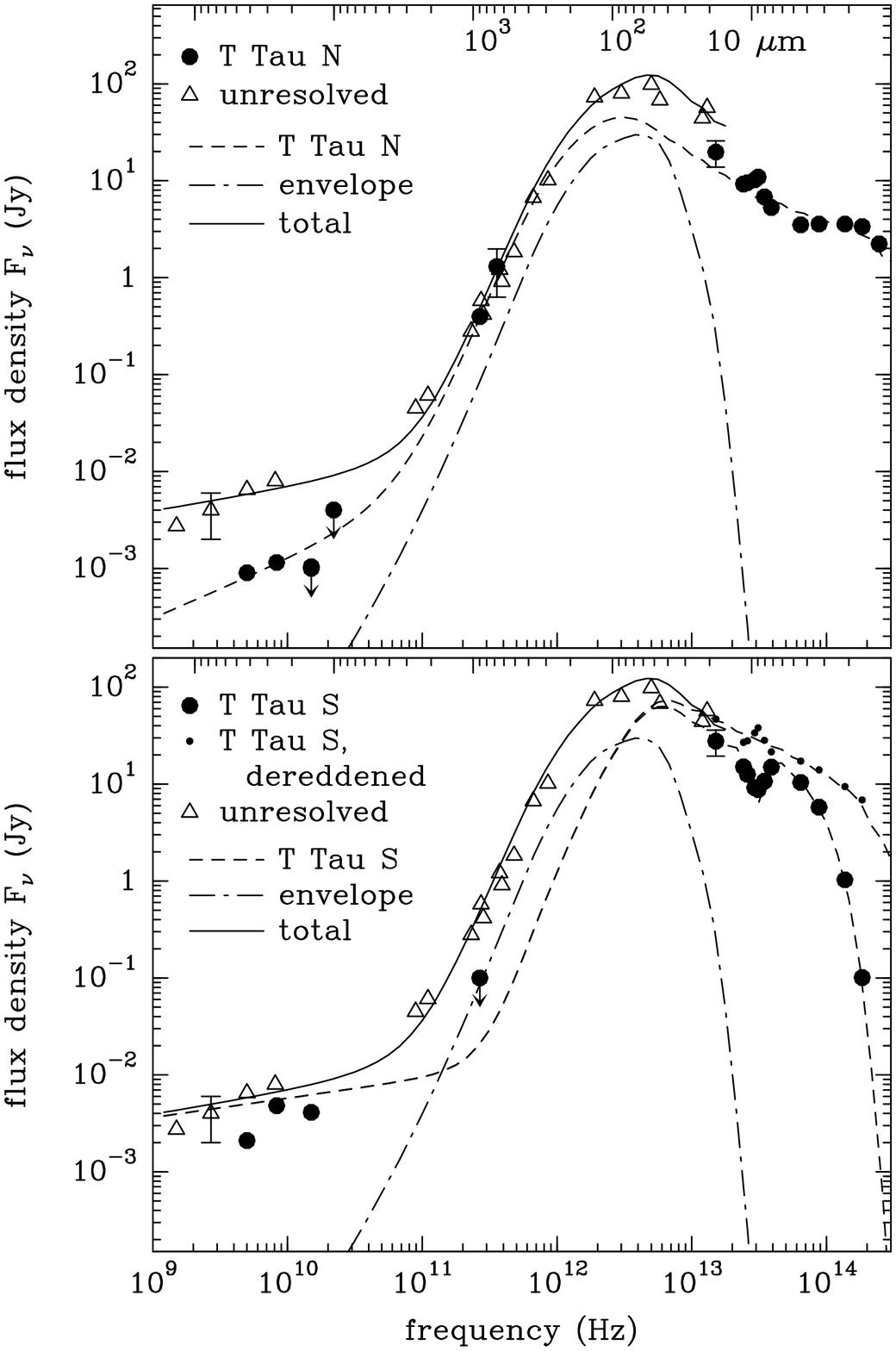,height=20truecm,rheight=11truecm}
}}
\vskip -1truecm

\end{document}